\documentstyle{mn}
\input epsf

\newcommand{\Msun}{\,M$_\odot$}
\title{Ages of globular clusters: a new approach}
\author[R. Jimenez et al.]
{Raul Jimenez,$^{1,2}$ Peter Thejll,$^{2}$ Uffe G. J\o rgensen,$^{2}$ 
 James MacDonald$^{3}$\cr and Bernard Pagel$^1$\\ 
$^1$NORDITA, Blegdamsvej 17, DK-2100 Copenhagen, Denmark \\ 
$^2$Niels Bohr Institute, Blegdamsvej 17, DK-2100 Copenhagen, Denmark  \\
$^3$Dept. of Physics and Astronomy, University of Delaware, Newark, DE 19716, USA }
\date{}

\begin{document}
\maketitle

\begin{abstract} We have applied a new method to analyze the horizontal 
branch (HB) morphology in relation to the distribution of stars near the 
red giant branch (RGB) tip for the globular clusters M22, M5, M68, M107, 
M72, M92, M3 and 47 Tuc. This new method permits determination of the 
cluster ages to greater accuracy than conventional isochrone fitting.  
Using the method in conjunction with our new high-quality photometric 
data for RGB and HB stars in the first five of these clusters, we discuss the 
origins of the spread in color on the HB and its relation to the 
`second parameter' problem. The oldest clusters in our 
sample are found to have relatively low ages ($13.5 \pm 2$ Gyr). 
 A 1$\sigma$ uncertainty 
in each of the parameters of mass and helium content combined 
with the effects of helium diffusion gives a lower limit for 
the age of the oldest clusters of 9.7 Gyr. \end{abstract} 


\section{Introduction}

Globular clusters (GC) probably contain the oldest identifiable stars and 
are therefore suitable for work on determining lower limits to the age of 
the Universe. Re-assessment of globular cluster ages is timely in view 
of recent ground-based and HST observations of Cepheid variables in 
Virgo cluster galaxies (Pierce et al 1994; Freedman et al 1994) which 
give a Hubble constant of $87 \pm 7$ and $80 \pm 17$ km s$^{-1}$ Mpc$^{-1}$ 
respectively. In the simplest inflationary Universe ($\Omega$ = 1, 
$\Lambda$ = 0) this corresponds to ages of $7.7 \pm 0.6$ and $8.7 \pm 
1.8$ Gyr, respectively, with a 50\% increase if $\Omega$ = 0. These age 
estimates are much lower than previously thought and in particular are 
lower than many estimates of the ages of the oldest globular clusters 
(Salaris, Chieffi and Straniero 1993, McClure et al. 1987). 
The era of CCD photometry has brought an impressive increase in the 
accuracy of the data available for GC analysis at the same time as rapid 
improvements in theoretical work on GCs. Deficiencies in the input 
physics combined with the uncertainties in cluster distances and 
interstellar reddening have made it difficult to determine globular 
cluster ages with an accuracy better than about 25\%.

The questions that we want to answer in this paper are:

\begin{itemize}

\item [i)] What is the origin of the spread in effective temperature 
along the HB?

\item [ii)] How is the morphology of the HB related to the morphology 
of the RGB tip? 

\item [iii)] Is it possible to constrain the mass loss efficiency and 
the mass of the RGB from the morphology of the HB?

\item [iv)] If so what is the age of the GCs and what are the uncertainties 
involved?

\item [v)] How is the `second parameter' related to the origin of the morphology of the HB?

\end{itemize}

Since the earliest theoretical stellar evolution sequences for low-mass 
stars, major efforts have been made to calculate globular cluster ages 
(VandenBerg 1988). The most popular method, isochrone fitting 
(Sandage 1982, VandenBerg 1983), is to fit the observed GC main-sequence 
with appropriate main sequence theoretical isochrones. The age of the 
cluster is gauged from the position of the observed main sequence turnoff 
relative to the isochrones. The most important ingredient of this method, 
as we see it, is the necessary assumption of a description of the 
efficiency of convective energy transport in modelling stellar structure 
and evolution -- often treated by a mixing length theory (e.g. B\"{o}hm--
Vitense 1958) -- and this assumption's influence on the structure of the 
star. The `mixing length parameter' ($\alpha$), which is one of several 
parameters that determine the convective efficiency (the others are often 
left at fixed canonical values), is often chosen {\it a priori}. Usually 
a value is chosen that is close to the value found in calibrations of 
solar-type models to the Sun's observed luminosity, radius and/or 
temperature (not all observables of the Sun are always used in the 
calibration, and unfortunately it is not uncommon to find fits in the 
literature that have been done using the solar models at a wrong age 
and/or metallicity -- see J\o rgensen (1991) for a more complete 
discussion). Although the Sun may not be similar to other low-mass stars 
in chemically different systems such as GCs, the assumption is, 
apparently, that we do at least know the Sun very well; this argument is 
not often touched upon, see VandenBerg (1983 section V) for one of the 
few published comments on this problem. The typical error in the age 
determination for the isochrone fitting method, excluding that due to the 
mixing-length parameter problem, is about 3 Gyr, and is mainly due to 
uncertainties in reddening, distance modulus and the position of the 
turnoff, as has been discussed by Chaboyer (1995).

In order to avoid some of these problems, Iben and Renzini (1984) 
presented another method to determine GC ages from a calibration of the 
magnitude difference between the horizontal branch (HB) and the 
main--sequence turnoff, at the color of the turnoff. 
The method relies on the 
fact that the luminosity of the ZAHB is probably not a function of age, 
while that of the turnoff obviously is. The main disadvantage of this 
method is that many GCs show either a very blue or a very red HB, and it is 
difficult to estimate with any certainty the location of the ZAHB at the color 
of the turnoff. This method also suffers from the difficulty of precisely 
locating the turnoff. A 0.3 mag. error (VandenBerg 1988) in the 
observed magnitude difference corresponds to an error of 5 Gyr in the age 
determination.  

A third approach to measure relative ages for star clusters with similar 
chemical compositions was developed by VandenBerg, Bolte and Stetson 
(1990). This method uses the fact that the color difference between the 
turnoff and the base of the giant branch, $\delta(B-V)$, decreases as the 
the cluster age increases. Therefore, aligning all the colour-magnitude 
diagrams (CMDs) and looking at the positions of the different RGBs will 
give an age estimate. The obvious limitation of this method is that it 
only applies to clusters with the same metallicity. Nevertheless, this 
method introduced an important procedure in the process of understanding 
GCs: relative morphological arguments.  Using relative morphological 
arguments the authors were able to give much more accurate relative ages 
than before. Unfortunately the same authors concluded that no relative 
morphological argument could be used to give an absolute age 
determination. We will show in this paper how this is not necessarily 
true and how the power of morphological constraints can lead to an 
accurate determination of GC ages. 

The above three methods have been applied to several globular
clusters.  The age found for clusters like M92 and M68 ranges from
values like 19 Gyr (Salaris, Chieffi and Straniero 1993), with the
$\alpha$--elements (O, Ne, Mg, Si, S, and Ca) enhanced relative to
iron, to 14 Gyr (VandenBerg 1988) for [O/Fe] enhanced models with the
ratios of the other $\alpha$--elements to iron kept at the solar value,
and a larger distance modulus. Therefore, the difference in the value
for the age is quite large (33\%).
It is also evident that an increasing number of `control-parameters'
(like oxygen enhancement, increased helium abundance, etc) affect the results
of such work.
For instance, Bergbusch \& VandenBerg (1992) tested
the effects on the age question of adding only oxygen. However, without
explicit calculations it is not clear how the non-solar abundance
ratios should be handled.

In order to illuminate this central question in cosmology and stellar 
evolution we have developed a new method to determine GC ages. The method 
is based on an accurate determination of the mass of the RGB stars and of 
$\alpha$ at the RGB, as well as quantitative determination of the mass 
loss along the RGB, which is constrained by the HB morphology. These 
theoretical improvements have been complemented with accurate multi-band 
photometry (UBVRIJHK) for five GCs that we have gathered during the past 
year. In addition, we have analyzed data from the literature for three 
other well studied clusters (M92, M3 and 47 Tuc). 

While the mixing length parameter $\alpha$ is but a crude approximation 
to stellar convection, we can show that its use is consistent and is 
meaningful in the relatively short-lived RGB stars in clusters, since at 
this stage of evolution of GC RGB stars the physical conditions of the 
stars are similar. 

Since GCs are observed to have extended HBs one infers that there is 
variation in the parameter or parameters that determine the colour of HB 
stars. The ratio of core mass to total stellar mass, ($q$), is one 
parameter that influences HB star colour, but metallicity also plays a 
role. As the stars evolve from the ZAHB, they may move along the HB and 
thus appear at a colour that is also a function of the time since arrival 
on the HB. It is well known that, of these factors, $q$ is the most 
important (Rood 1973). In the ratio $q$, the core mass for stars recently 
arrived from the RGB tip (i.e. within a few HB star lifetimes - a few 
times $10^{8}$ years), is a narrowly constrained number near 0.5 
$M_{\odot}$ due to the physics of the He core flash in low mass stars. 
The only variable that is left for producing variations in $q$, and thus 
in HB color, is the total mass or, since the core mass is more or less 
fixed, the mass of the envelope left on top of the core following the RGB 
evolution. The early work by Rood (1973) made clear the necessity for a 
spread in the mass to explain the morphology of the HB 
(Lee, Demarque \& Zinn 1994), but there was no 
attempt to do a quantitative analysis and link its properties with the 
rest of the HR-diagram. J\o rgensen \& Thejll (1993) investigated 
quantitatively which parameters can model the HB consistently with the 
morphology of the RGB. They concluded that only two possible scenarios 
could account for the HB morphology: stochastic variations in the mass 
loss efficiency and/or a delayed helium core flash due, for instance, to 
rotation. With our new observational material and new theoretical tools, 
we now conclude that the morphology of the HB can only be explained by 
stochastic variations in the mass loss efficiency. Also we put strong 
limits on the variations in the mass loss efficiency by analysing the 
morphology at the RGB tip. 

In this paper we present new observations and theoretical tools that 
bring clues about three important subjects in GC studies: the physical 
origin of the spread in temperature on the HB, the `second parameter' 
problem and the age of the GCs. In 
sections 2 and 3, we describe the observations and data reductions. In 
section 4, we describe the theoretical tools used to analyse HB morphology 
in GCs and red giant branch tip (RGT) morphology. In section 5, we 
present the analysis of the HB morphology in terms of differential mass 
loss processes. In section 6, we use the previous tools to analyse the set 
of GCs and calculate their ages. We finish in section 7 with a critical 
comparison with previous studies from the literature.

\section{Observations and data reductions}

We obtained UBVRIJHK photometry of the GCs M22, M107, M72, M5, and M68 
in June 1993 at La Silla (Chile) using the Danish 1.5m telescope and the 
ESO 2.2m telescope. Here we concentrate on the optical photometry --
the IR photometry will be described in a forthcoming paper. 

The observations at the Danish 1.5m telescope were taken with a 1024 
$\times$ 1024 pixels CCD. Dark, bias and flat frames were taken in order 
to remove the instrumental signature from the data. The flat field 
frames were observed both on the sky during twilight and on the dome. The 
clusters were covered in each filter with one frame, but several frames 
were taken of the same cluster and the same filter in order to improve 
the signal-to-noise ratio. Typical exposure times were 60 s to 200 s for 
the V filter, and about 1800 s for the U filter. For the flat field 
twilight exposures we used a range of 30 s for the B,V,R and I filters 
to 200 s for the U filter. Since we were only interested in 
investigating the giants and HB stars, we did not perform photometry 
deep enough to reach the main sequence turnoff in most cases.

All data reductions were carried out with utilities in the IRAF package. 
All science frames were corrected for dark current, bias level and  pixel to pixel
variations in response to illumination. Standard stars with a wide range 
in colour, selected from the list of Landolt (1992), were observed 
through the night at various air masses.

Photometry of cluster stars was reduced using DAOPHOT II (Stetson 1987). 
The point spread function was determined from 10 bright, isolated stars 
in each frame. A constant and a linear model for the PSF were used to 
extract magnitudes; since both methods lead to the same result, a 
constant PSF was chosen in the reductions. Averaging of the 
approximately 20 available frames for each filter was performed to 
obtain the best possible signal. 

In order to improve the accuracy of the photometry to the statistical 
limit and to obtain the best possible results for a morphological 
analysis, we used an additional algorithm to decrease the errors in the 
magnitudes. This is based on the idea that while all stars in a single 
frame may be well calibrated relative to the standards, the standard 
star fields and the cluster frames were necessarily taken at slightly 
different times and in different directions on the sky. Small, otherwise 
undetectable atmospheric disturbances, variations in the atmospheric 
water content, turbulence, small wings in the PSF, etc., could 
therefore offset each cluster frame from the others which will result in 
a larger error than that due to the calibration process itself. We 
selected in one frame around 100 of the most well behaved stars, i.e. 
those that are reasonably isolated and far away from the edge of the 
frame and which show low scatter in the color-magnitude diagram (CMD) -- 
such as a section of the RGB between the top of the sub-giant branch and 
the HB.  Using the coordinates of these stars, we found them in each of 
the frames for the relevant filter.  In each frame, we then calculated 
the 100-star average of the selected stars and then expressed the 
magnitudes of all stars in each frame relative to the 100-star average.  
This step removes small offsets in the photometry since each frame is 
probably affected uniformly across the field by any thin cirrus or dust. 
By averaging these frames we can then reduce the statistical errors. It 
then remains to calibrate these `atmospheric disturbance--free' frames 
onto the standard star system of photometry. This step is done by 
shifting the `atmospheric disturbance--free' frames back to the 
photometric values present in the calibrated and averaged `possibly 
slightly atmospheric disturbed' frames of the previous paragraph. If 
the effects of the postulated atmospheric disturbances are normally 
distributed then the above process will lead to photometry with no worse 
offsets in magnitudes than those present in the `possibly slightly 
atmospheric disturbed' frames and with photometric errors on each star 
substantially lower.  As shown in Figure 2 for the cluster M68, the 
improvement in the photometric errors is quite significant. In Figures 
2 to 6  we show the HR-diagrams for the five GCs observed. 

\begin{figure}
\centering
\leavevmode
\epsfxsize=1.0
\columnwidth
\epsfbox{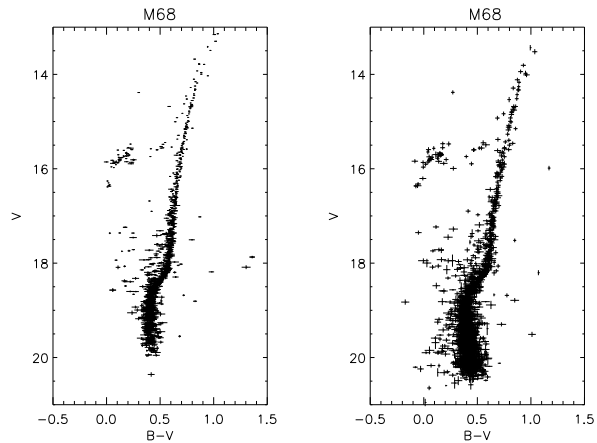}
\caption[]{The figure shows the errors in the photometry of the
globular cluster M68 using two methods to reduce the data.  The panel
to the right shows the standard method of averaging, while the panel to
the left shows the improvements in photometry due to the additional
reduction step we have introduced -- see section~2. The brightest stars
at V=12.5$^m$ have not been included here owing to overexposure  (see
McClure et al 1987, where the brightest stars were taken from
photographic photometry).} 
\end{figure}

\begin{figure}
\centering
\leavevmode
\epsfxsize=1.0
\columnwidth
\epsfbox{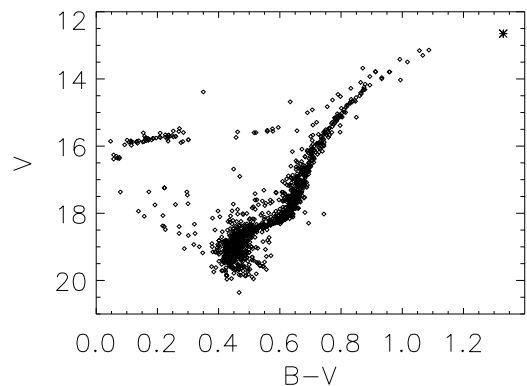}
\caption[]{HR-diagram for the globular cluster M68. The size of the symbols is 
about that of the errors. There is a clear split between the RGB and 
the AGB. The  ``star'' shows the position of the two reddest stars in the
cluster measured photographically by Harris (1975).}
\end{figure}

\begin{figure}
\centering
\leavevmode
\epsfxsize=1.0
\columnwidth
\epsfbox{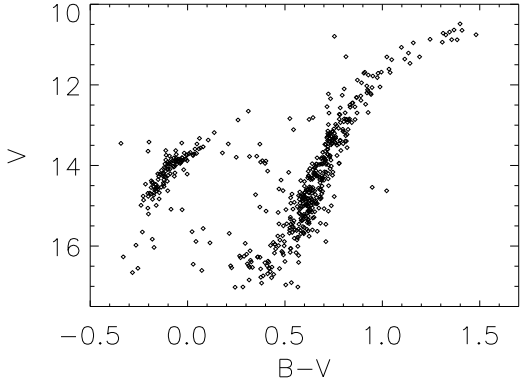}
\caption[]{Colour--magnitude diagram for the globular cluster M22. The cluster shows a very blue HB and a broad RGB. The thickness of the 
RGB is bigger than the photometric errors. In the text we show how this 
intrinsic broadening of the RGB and AGB can be due to metallicity variations 
in the GC.}
\end{figure}
 
\begin{figure}
\centering
\leavevmode
\epsfxsize=1.0
\columnwidth
\epsfbox{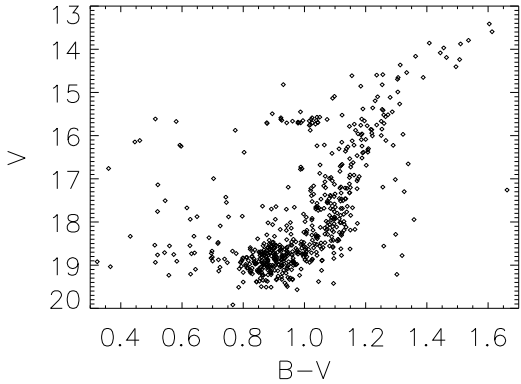}
\caption[]{The GC M107 has a very broad RGB. The errors in the photometry 
are much smaller than the thickness of the RGB -- the size of the symbols 
is approximately that of the errors. It is important to be aware of
the field contamination in this cluster.
} 
\end{figure}
 
\begin{figure}
\centering
\leavevmode
\epsfxsize=1.0
\columnwidth
\epsfbox{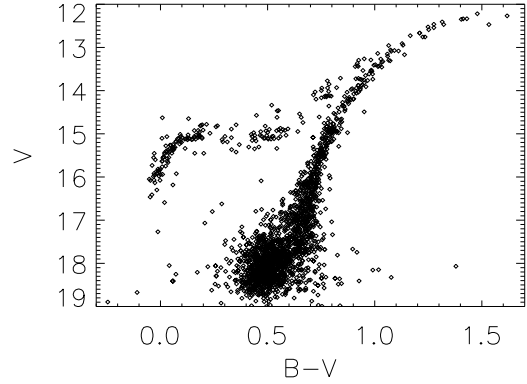}
\caption[]{M5 is another globular cluster with a clear split between the RGB 
and the AGB. It is easy to distinguish AGB and RGB stars in the diagram.}
\end{figure}
 
\begin{figure}
\centering
\leavevmode
\epsfxsize=1.0
\columnwidth
\epsfbox{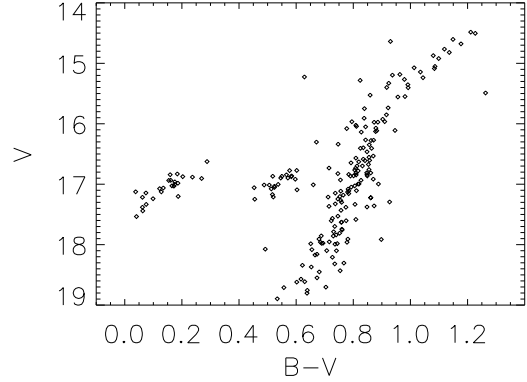}
\caption[]{M72 is one of the least studied GCs in the literature. Here we 
present for the first time CCD data for this cluster. The HB shows a 
well defined RR Lyrae gap.  
}
\end{figure}

\begin{figure}
\centering
\leavevmode
\epsfxsize=1.0
\columnwidth
\epsfbox{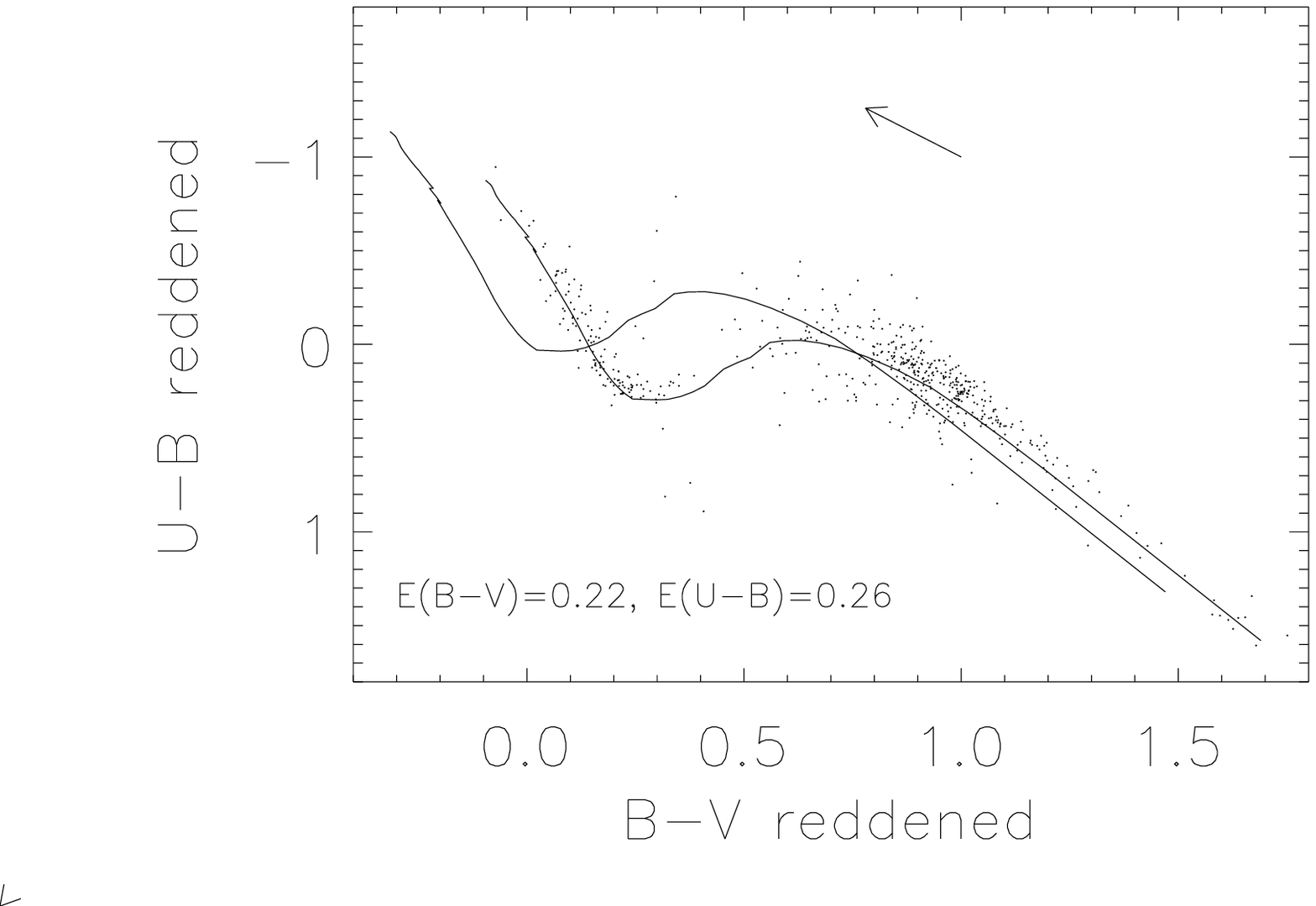}
\caption{The two color diagram $(U-B)$ vs. $(B-V)$ for M22. We show the fit 
with an arbitrary reddening law -- $E(U-B)/E(B-V)$ 
taken as a free parameter -- to the data using Kurucz (1993) stellar 
atmosphere models. Also we plot the de-reddening arrow that shows the 
direction of the applied correction. In addition to this we have 
plotted the positions of Kurucz models with and without reddening correction 
applied.} 
\end{figure}

To make theoretical predictions, we need to transform the observed $(B-
V)$ vs. $ V$ diagram to the theoretical $L$ vs. $T_{\rm eff}$ diagram. To 
do this it is necessary to know several parameters, such as the distance 
modulus $(m-M)_V$, the reddening $E(B-V)$, the metallicity Z, the helium 
content Y and the theoretical transformation from $(B-V)$ to $T_{\rm 
eff}$. The distance modulus $(m-M)_V$ is taken from the several values 
listed in the literature. In order to obtain good average values for 
distance modulus, metallicity and reddening, we formed  weighted averages 
of values taken from the literature, with the weights inversely 
proportional to the square of the statistical errors. It is important to 
note that in some cases the number of available measurements is not very 
large (3 or so), and that in such cases the statistical procedure was not 
employed. We have also calculated the reddening by comparing the $(B-V)$ 
vs. $(U-B)$ diagram from our data with theoretical colours from Kuruzc 
(1993) using a least squares method. In Table 1 we show the range of metallicity data, distance modulus, reddening taken from the literature 
and our adopted average values, we also include the HB colour range. 


\begin{table*}
\begin{tabular}{ccccc}
\hline
& M68 & M5 & M72 & M107 \\
$[Fe/H]$ & $-2.1$ to $-2.0$ & $-1.0$ to $-1.58$ & $-1.5$ & $-1.3$ to $-0.7$ \\
$[Fe/H]$(a)& $-2.0$ & $-1.3$ & $-1.5$ & $-0.9$ \\ 
$(m-M)_V$&15.01 to 15.30&14.51 to 14.55&16.30 to 16.50&14.97 to 15.03 \\
$(m-M)_V$(a)&15.25&14.53&16.50&14.97 \\
$E(B-V)$&0.03 to 0.07&0.01 to 0.05&0.03 to 0.07&0.23 to 0.31 \\
$E(B-V)$(a)&0.07&0.02&0.07&0.23 \\
HB $(B-V)$&0.05 to 0.60&$-$0.05 to 0.6&0.0 to 0.6&0.85 to 1.1 \\ 
\hline
& M22 & M92 & M3 & 47 Tuc \\
$[Fe/H]$&$-1.5$ to $-2.0$&$-2.19$ to $-2.1$&$-1.69$ to $-1.57$&$-0.70$ to $-0.46$ \\
$[Fe/H]$(a)& $-1.8$ & $-2.1$ & $-1.6$ & $-0.6$ \\
$(m-M)_V$&12.70 to 13.60&14.45 to 14.51&14.97 to 15.01&13.44 to 13.46 \\
$(m-M)_V$(a)&13.50&14.45&15.01&13.46 \\
$E(B-V)$&0.24 to 0.37&0.00 to 0.01&0.01 to 0.05&0.03 to 0.04 \\
$E(B-V)$(a)&0.22&0.00&0.02&0.04 \\
HB $(B-V)$&$-0.25$ to 0.2&$-$0.15 to 0.53&0.05 to 0.6&0.7 to 0.9 \\
\hline
\end{tabular}
\caption{The table shows the range of values for the metallicity,
distance modulus and reddening taken from the literature and the
adopted value in this paper. We also show the observed range in 
reddening-corrected $(B-V)$
colour of the HB for the set of GCs.
} 
\end{table*}

We use $Y=0.24 \pm 0.01$ for all GCs based on standard arguments
related to big bang nucleosynthesis and chemical evolution (Dorman,
VandenBerg \& Laskarides 1989, Pagel 1992). The bolometric correction
and $T_{\rm eff}$ are calculated from the latest version of Kurucz
stellar atmosphere models (Kurucz 1993). In Figures 2 to 6 we show the HR 
diagrams for the five GCs observed, corrected for reddening.

\section{Globular cluster reddening, distance modulus, metallicity, and
field star contamination}

\subsection{M68}
The galactic globular cluster M68 ($l=299.6,b=36.0$) is a metal-poor halo 
globular cluster that, due to its relative openness and low reddening, is 
suitable for resolution of the center of the cluster. The most relevant 
recent studies are those by Walker (1994), Alcaino et al. (1990) and 
McClure et al. (1987).

From the several methods available to calculate the reddening in the
cluster, only one was possible using the data that we collected; the 
two-color diagram method in which the reddening is estimated from the 
offset needed to shift the observed colour-colour diagram onto the 
theoretical two colour diagram. A major problem is that the stellar 
atmosphere models used to construct the theoretical diagram do not 
describe the real stars equally well at all positions in the diagram.  
Another problem is that for the occasional cluster that lacks a well-
defined HB, it is uncertain how to match the theoretical 
colors to the observed ones since only the RGB and whatever part of the 
sub-giant branch and MS that have been observed are available, and the 
RGB can be the part of the diagram where theoretical models are the 
least realistic, mainly due to problems of including enough 
opacities at cool temperatures. Fortunately, for M68 the reddening is 
not large since the cluster is far from the galactic disk. Our fit to 
the observed two-color diagram with a theoretical grid of stellar 
atmosphere models (Kurucz 1993) gives a reddening of $E(B-V)=0.06$. For 
comparison, McClure et al. (1987) and Walker (1994) found $E(B-V) = 0.07 
\pm 0.01$, which is in good agreement with our value.

We adopted a value of $[Fe/H] = -2.0 \pm 0.10$. 

Another source of error in the cluster color-magnitude diagrams is  
contamination by field stars. We estimate the severity of this 
effect by calculating for each cluster the number of expected field 
stars in the area covered by the CCD given the size of the CCD field and 
its galactic latitude, we have used for this purpose the model by Bachall \& Soneira (1981). For M68, the number of expected field stars is 
around one for magnitudes brighter than 16, and so field contamination 
is not a serious problem. 

To transform the observed HR diagram to the theoretical 
luminosity--effective temperature diagram, 
we need to know the cluster's distance 
modulus. From the values given in Walker (1994), Alcaino et al. (1990) 
and McClure et al. (1987) for M68, we have adopted an apparent distance modulus of 
$(m-M)_{V}=15.2 \pm 0.1$. 

\subsection{M22}
 
M22 is a bright globular cluster that lies in the disk of the Galaxy 
($l=9.9, b=-7.6$) and the problem of field contamination is therefore crucial.  
The number of field stars brighter than magnitude 16 is estimated to be 
about fifteen. The most recent studies of this cluster are those of 
Samus et al.  (1995), Peterson \& Cudworth (1994), Cudworth (1986) and 
Alcaino and Liller (1983).  

Determination of the reddening for this cluster is still an open 
question. The published values have a large spread, ranging from $E(B-
V)=0.24$ (Crawford and Barnes 1975) to $E(B-V)=0.37$ (Zinn 1980). Because 
this cluster has a well defined blue HB, we have used the two colour 
diagram method to find the reddening. We used two approaches. The first 
approach was to use a standard reddening law. The 
second approach was to let $E(U-B)/E(B-V)$ be a free parameter. 
 We found 
that the second approach with $ E(U-B)/E(B-V)=1.2$ produced a much better 
fit, giving a reddening of $E(B-V)=0.22$ and $E(U-B)=0.26$. We have 
therefore adopted these results for the reddening. In Fig. 7 we show the 
fit to the observed data and also the de-reddening arrow.

M22 was one of the first globular clusters for which a color-magnitude 
diagram was obtained (Arp \& Melbourne 1959).  The cluster has a low 
metal abundance, but the value is uncertain due to the high field star 
contamination. Several spectroscopic studies of M22 exist. Peterson 
(1980) derived metallicities ranging from $[Fe/H]=-1.62$ to $-2.18$, from 
echelle spectra of four giant stars. Cohen (1981) carried out a 
detailed abundance analysis using high dispersion echelle spectra of
three red giant stars and determined that they were chemically identical,
with $[Fe/H]=-1.78$. Gratton (1982), also from high dispersion echelle 
spectra of three stars, found no indications of inhomogeneities for any 
element, deducing $[Fe/H]=-1.94$.  Pilachowski et al. (1982) made an 
echelle spectral analysis of six stars near the red giant tip and derived 
iron abundances ranging from $[Fe/H]=-1.4$ to $-1.9$. Our weighted 
average value for the metallicity is $[Fe/H]=-1.8$ with an uncertainty of 
0.25. The spread in metallicity seems to be real and could be the 
reason why the RGB is intrinsically broadened in this cluster. From Fig. 
3, it can be seen that the photometric errors for individual RGB stars are 
much smaller than the width of the RGB. We have computed the expected 
spread in the RGB if the metallicity ranged from $[Fe/H]=-1.5$ to 
$[Fe/H]=-2.0$. At a luminosity of $log\,L/L_{\odot}=2.8$, the spread 
would be 0.015 dex in $log \, T_{\rm eff}$, which corresponds to a 
spread of 0.2 in $(B-V)$. This is roughly the spread observed in the RGB 
of M22, and so we conclude that it is likely that the RGB in M22 is 
broadened by an intrinsic variation of the metal content in the cluster. 
 The possibility of differential reddening 
(that would produce the same spread as the metallicity does)  
along the cluster cannot be completely ruled out -- 
but it is very unlikely because  stars 
with different metallicity in the cluster scatter all over it. 

Due to the uncertainties in the reddening, an estimation of the distance 
modulus is difficult. We have adopted a value $(m-M)_{V}=12.8\pm 0.3$. 

\subsection{M72}
 
The galactic globular cluster M72 
($l=35.1,b=-32.7$) 
is not a well studied cluster. It has 
low metallicity and low reddening, $E(B-V)=0.07$ (Dickens 1972). The 
cluster is relatively open and therefore suitable for resolving the core. 

Even though the HB is not as well defined as in M22, we have, as in the 
previous cases, used the two colour $(B-V)\, vs.\, (U-B)$ diagram to 
determine the reddening. We find $E(B-V)=0.07$ using a standard reddening 
law, in perfect agreement with the value found by Dickens. The 
metallicity for the cluster, $[Fe/H]=-1.4$, has been taken from Harris \& 
Racine (1979), and the distance modulus, $(m-M)_{V}=16.3 \pm 0.1$, has 
been taken from Dickens (1972). Due to the characteristics of M72, the 
uncertainties due to reddening and metallicity are not nearly as great as 
for M22. 

Field contamination is more severe in this case than in M68, but even so, 
the number of expected field stars is only five for magnitudes brighter 
than 17. 

\subsection{M5}
 
M5 ($l=3.9, b=46.8$) has been the subject of many studies, 
most recently by Peterson 
(1979), Zinn (1980) and Buonanno, Corsi \& Fusi Pecci (1981). It has low 
reddening and intermediate metallicity (Peterson 1979, Searle and Zinn 
1978). From the two-colour diagram method using a standard reddening 
law, we determine $E(B-V)=0.02$, in perfect agreement with the value 
from Buonanno et al. (1981). The value for the metallicity given by 
Buonanno et al. (1981) corresponds to $[Fe/H]=-1.3$, but a considerable 
range of values exists in the literature, from $[Fe/H]=-1.0$ (Butler 
1975) to $-1.58$ (Zinn 1980). We have adopted $[Fe/H]=-1.2$, based on 
the weighted average of the literature values.

The well populated RGB is not intrinsically broad and our data allow a 
clear separation between the RGB and AGB stars.  

\subsection{M107}

This cluster ($l=3.4,b=23.0$) has a low central concentration. 
The two most recent 
papers on M107 are Ferraro et al.(1991) and Zinn (1985). 
It is metal-rich, with [Fe/H]$=-0.9$ (Harris and Racine 1979). 
The reddening is 
quite uncertain, and again the RGB appears to be extremely broad, more 
than the intrinsic errors in the photometry. 

The literature value for the reddening is $E(B-V)=0.31$ (Zinn 1985).  
 Using the 
two colour diagram method with a standard reddding law, we find $E(B-
V)=0.28$. However the alternative reddening law with $E(U-B)/E(B-V)$ a 
free parameter gives a better fit. We find $E(U-B)/E(B-V)=1.13$,  
 $E(B-V)=0.23$ and $E(U-B)=0.26$.

We have studied, as in M22, whether the reason for the broadened RGB 
could be different metallicities in the cluster. Again we have reached 
the conclusion that a range in metallicity of 0.5 dex could explain 
the RGB colour spread.

The field contamination for this cluster is not high. The number of 
expected stars for magnitudes brighter than 16 in $V$ is around 10. 

\subsection{Other globular clusters}
We have added to our sample three other GCs: M92, M3, and 47 Tuc. M92 
has similar metallicity to M68 ([Fe/H]$=-2.1$ (VandenBerg 1983)).  Data
for the RGB in M92 is taken from the compilation in VandenBerg (1983)
which relies on data from Sandage\&Walker (1966) and Sandage (1969,
1970),  and for the HB the colour of the reddest HB star is set
by the CMDs in Sandage\&Walker (1966), Buonanno et al. (1983, 1985), 
Rees (1992) and Montgomery\&Janes (1994). The presence of HB stars on the
red side of the RR Lyrae gap is in general confirmed by the values
reported for (B-R)/(B+R) or B/(B+R), where B and R are the numbers of
non-variable stars on the blue and red side of the RR Lyrae gap
respectively, by Madore (1980) and Lee, Demarque\& Zinn (1990) as well as
the mass-distribution for the non-variable stars on the HB presented by
Crocker, Rood\&O'Connel (1988).  47 Tuc has been analysed in detail by
Frogel, Persson and Cohen (1981) and by Dorman, VandenBerg and
Laskarides (1989). We have adopted a value of [Fe/H]$ = -0.6 $ for this
cluster. For M3, using photometric data from Buonanno et al (1986), we have
adopted [Fe/H]$ = -1.6$ (Harris \& Racine 1979).

\section{Stellar evolution models}

The theoretical analysis of the GCs has been carried out with two 
theoretical tools: a classical set of stellar evolution tracks for low 
mass stars which we have calculated for this purpose using up-to-date 
input physics, and a semi--analytical stellar evolution code based on 
the evolution models. 

Since the first grid of stellar evolution sequences for low mass stars 
from the main sequence to the RGT (Sweigart \& Gross 1978), a large 
effort has been made to accurately describe the evolution of low mass 
stars along the RGB. Many grids of stellar evolution sequences have been 
published since then:  VandenBerg \& Bell (1985), VandenBerg (1992), 
Charbonnel et al. 1993, Fagotto et al. (1994).  Why then the necessity for 
a new grid? We aimed to study the effects of variations in several 
parameters (mixing length ratio $\alpha$, mass-loss efficiency $\eta$ as 
parameterised in Reimers' mass-loss `law' (Reimers 1975), total mass $M$, 
 helium abundance $Y$ and 
metallicity $Z$) on the CMD morphology. This requires a large internally 
consistent grid spanning the relevant parameters; such a grid does not 
exist in the literature. We have, therefore, 
calculated 130 stellar evolution 
sequences from the contracting Hayashi phase to the RGT. The set of 
parameters chosen for the grid of models is given in Table 2. 

\begin{table}
\begin{center}
\begin{tabular}{|c|c|c|c|}
\hline
$M/M_{\odot}$ & $Z$ & $\alpha$ & $Y$ \\
\hline
0.55 & 0.0002 & 1.00 & 0.24 \\
0.60 & 0.0005 & 1.25 & 0.28 \\
0.65 & 0.001 & 1.50 &  \\
0.70 & 0.004 & 1.75 &  \\
0.75 & & 2.00 &  \\
0.80 & & &  \\
0.85 & & &  \\
0.90 & & &  \\
0.95 & & &  \\
1.00 & & &  \\
\hline
\end{tabular}
\end{center}
\caption{The table shows the stellar parameters used in the grid of 
stellar evolution models for the analysis of the GCs.}
\end{table}

The stellar evolution code and the grid of tracks is described in 
detail in Jimenez \& MacDonald 1995. A detailed 
description of the input physics can be found in Jimenez et al. (1995). 

\begin{figure}
\centering
\leavevmode
\epsfxsize=1.0
\columnwidth
\epsfbox{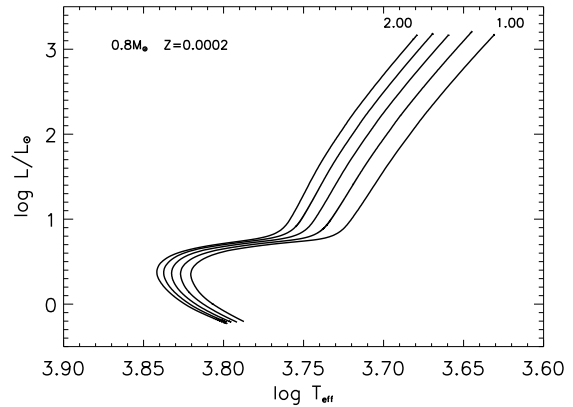}
\caption[]{Effect on the RGB and main sequence turnoff of different values
of $\alpha$. The RGB is more
sensitive to changes in $\alpha$ than the turnoff region.
}
\end{figure}

\begin{figure}
\centering
\leavevmode
\epsfxsize=1.0
\columnwidth
\epsfbox{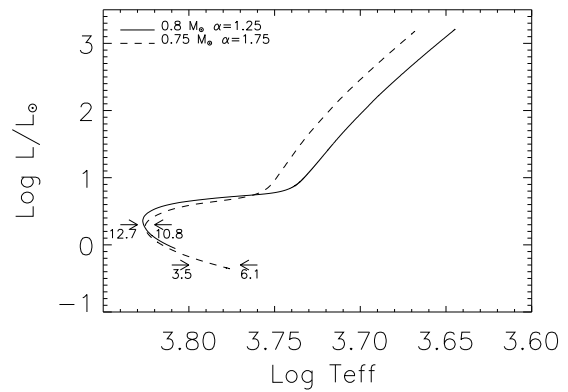}
\caption[]{Two tracks with slightly different mass and $\alpha$ show zero
sensitivity at the turnoff point. This could lead to a wrong determination
of the age. The arrows show the age in Gyr. for both tracks. The arrow
on the right refers to the 0.8 $M_{\odot}$ track, and the arrow on the left
to the 0.75 $M_{\odot}$ track.}
\end{figure}

In Fig. 8 we show how the RGB is more sensitive to $\alpha$ than the 
turnoff. The effective temperature of stellar models  is at least 
four times more sensitive to $\alpha$ when they are on the RGB than when they are at the 
main sequence turnoff.
The effect of a wrong choice of $\alpha$ in the turnoff region is shown 
in Fig. 9. In the figure we show how two stars with different masses 
(0.75 and 0.80 $M_{\odot}$, respectively) and different values of 
$\alpha$, lie in the same position at the turnoff point. In the two 
tracks calculated $\alpha$ took values 1.25 and 1.75 respectively. From 
the figure we observe that the difference in log $T_{\rm eff}$ and in 
luminosity at the turnoff 
point between both tracks is 0.01 dex, and  0.2 mag, respectively -- 
quite inside the typical 
observational error (see Fig. 2). On the other hand the RGBs of both 
clusters present a clear split of 312 K. Therefore the RGB seems to be a 
safer place to avoid ambiguities in the determination of $\alpha$. If an 
isochrone is calculated from these two tracks in order to make a turnoff 
point fit, an error of 0.05 $M_{\odot}$ could be made, which leads to an error 
in the age estimate of 2 Gyr. 

The interesting feature of the turnoff point is that it is sensitive to 
the mass of the stars; therefore, in principle, it should be efficient 
to determine the age -- a star spends 90\% of its life on the main 
sequence -- of the stars in the GC using this technique. However the 
location of the turnoff is not observationally well defined. This means 
that, in fitting an isochrone to the turnoff point, one can choose the 
wrong value for $\alpha$ and hence make an error in the mass for the 
stars. One of the main sources of error when 
using isochrone fitting is the bad definition of the turnoff point.
It is not yet clear whether the observational spread at the turnoff point is an 
intrinsic one, or is due only to observational errors. 

The other tool that we have used is a semi-analytical stellar evolution 
code. A complete description of the code can be found in J\o rgensen 
(1991), J\o rgensen \& Thejll (1993) and Jimenez et al. (1995).  Here we will 
just give a brief description. 

The semi--analytical method is suitable for analysis of stellar evolution 
on the red giant branch and on the asymptotic giant branch, with complex 
mass-loss scenarios included. Mass loss on the RGB is determined from
Reimers' formula (Reimers 1975) with the mass loss efficiency parameter
$\eta$ described by a realistic distribution function based on direct
observations of mass loss. This method, which we shall refer to as
synthetic stellar evolution (SSE), relies on matching 
observational data of globular cluster red giant branch stars
to theoretical results obtained by
interpolation in grids of stellar evolution tracks.  The key points of
the synthetic method are that detailed stellar evolution models are
used for the interpolation, and that the parameters in stellar
evolution models, i.e., the mass-loss efficiency parameter in the
Reimers formula and the mixing length parameter ($\alpha$), are
determined by matching the observations of the RGB as well as the HB.
This assures that the physics in the SSE behaves correctly in a
relative sense and is calibrated to reality which ensures the right
absolute behaviour.

\begin{figure}
\centering
\leavevmode
\epsfxsize=1.0
\columnwidth
\epsfbox{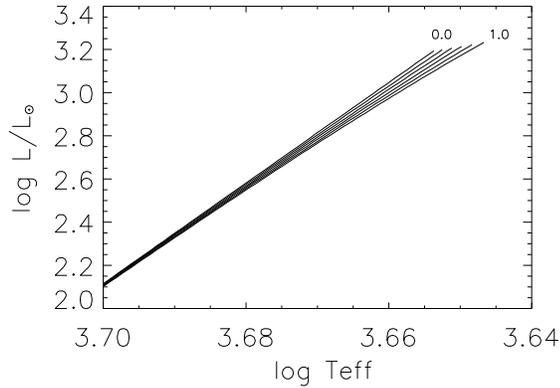}
\caption[]{The effect of mass loss at the RGT. The mass loss is
only important at the RGT, at the base all the tracks are similar. The
$\eta$ parameter has taken values from 0.0 to 1.0 in steps of 0.2. 
The tracks have been
calculated for a model of $0.8 M_{\odot}, \alpha=1.4$ and $Z=0.0002$.  }
\end{figure}

In brief, the SSE works in the following way: 

\begin{itemize}

\item[1] The RGB part of evolutionary tracks in a given grid is fitted 
with analytic formulas which express the relation between $L$, $T_{\rm 
eff}$, $M$, $M_{\rm c}$ (core-mass), $Z_{0}$ (metallicity, scaled 
according to [Fe/H]), $Y$ (helium abundance) and $\alpha$. The fitting 
formulas for the grid of stellar evolution models (Jimenez \& MacDonald 
1995) are: 

\begin{eqnarray*}
\log\,L\,=\,4.909 + 0.032\,\log\,Z_0 - 0.010\,{(\log\,Z_0)^2}  \\ 
+ 2.967\,\log\,(M_{c}) - 0.129\,\log\,Z_0\,\log\,(M_{c})  \\ 
 -  3.480\,{(\log\,(M_{c}))^2}
\end{eqnarray*}
\begin{eqnarray*}
\log\,T_{eff}\,=\,3.569 + 0.0640\,M + 0.0126\,{M^2} \\ 
- 0.128\,\log\,L - 0.145\,\log\,Z_0 + 0.0148\,M\,\log\,Z_0 \\ 
- 0.021\,\log\,L\,\log\,Z_0 - 0.0250\,{(\log\,Z_0)^2} \\
+ 0.094\,\log\,{\alpha} - 0.027\,M\,\log\,{\alpha} \\ 
+ 0.045\,\log\,L\,\log\,{\alpha} - 0.021\,{(\log\,{\alpha})^2}
\end{eqnarray*}
\begin{eqnarray*}
M_{c}^{RGT}\,=\,0.456 - 0.056\,M + 0.024\,{M^2} \\ 
- 0.016\,\log\,Z_0 + 0.002\,M\,\log\,Z_0
\end{eqnarray*}

where $Z_0$ is the solar heavy--element mass fraction scaled by [Fe/H],
log $Z_0=$ log$ Z_{\odot} +$ [Fe/H].  The goodness of fit of these
equations is evaluated across the grid of points used and is always so good
that errors in the observations exceed the error due to fitting.

\item[2] The metallicity is taken from model atmosphere analyses of 
observed spectra in the literature. The helium abundance $Y$ is set to 
0.24 from big bang nucleosynthesis arguments (Pagel 1992). 

\item[3] With the given $Z_0$ and $Y$, it was possible to fit all the
studied GCs (this paper, J{\o}rgensen \& Thejll 1993) to the analytical
expressions of point 1 (or the original tracks) by use of a nominal
value of the mass on the order of 0.8 \Msun and a single value of
$\alpha$. 

\item[4] The detailed value of M and the average value of $\eta$ are
then determined by requiring agreement between the observed HB mass
distribution (mean and dispersion to the red) and the modelled mass
distribution calculated using numerical integration of the mass loss
along the RGT.
A very fast and
accurate numerical computation of the evolution along the RGB is now
performed by taking advantage of the expressions of point 1. The
addition of mass to the core during a given time step in the
integration along the RGB is determined on the basis of the
instantaneous luminosity, the known energy generation rate, and the
length of the time step.  The mass of the core (M$_{\rm c}$) at the end
of a time step determines the new value of L according to the formulas
in point 1. The total stellar mass at the end of each time step is
calculated as the mass at the beginning of the time step minus the mass
loss rate times the length of the time step.  The evolution of the
synthetic track is stopped when M$_c$ reaches the value determined in
step 1 for the He-core flash.

\end{itemize}

In a recent paper, we have demonstrated the correctness of the SSE 
(Jimenez et al. 1995). In particular we have shown how the SSE is 
accurate when calculating scenarios with mass loss and how it can be used 
to model the evolution along the RGB. We showed that $T_{\rm eff}$ and 
luminosity at the RGT are so relatively insensitive to the parameters of 
the core that the SSE reaches the same values as do the full self--
consistent numerical solutions, even though for evolutionary 
phases with such rapid 
mass--loss, the full information of the mass loss does not `reach' the 
core before the core-flash. 

An important item when analysing GCs is the abundance of the $\alpha$--
elements. Many studies have found that these elements seem to be 
enhanced relative to Fe, compared to solar composition. Various 
observations have been performed in GCs in order to determine the 
abundance of individual elements in their RGB stars. Pilachowski, 
Wallerstein \& Leep (1980) studied M3 and M5 and found [O/Fe] $\sim$ 
0.3, [Si/Fe] $\sim$ 0.3. Similar results were found by Cohen (1978) in 
M3 and M13.  Gratton, Quarta \& Ortolani (1986) found the same results 
for all the $\alpha$--elements in 47 Tuc, M4, M5, NGC 6752 and M71. In 
the same vein are the results of Gratton \& Ortolani (1989) for NGC 
1904, NGC 3201, NGC 4590, NGC 4833, NGC 6254, NGC 6397, and NGC 6656, 
and Peterson, Kurucz \& Carney (1990) for M92. Observations for stars in 
the halo also point in the direction that all the $\alpha$--elements are 
in fact enhanced relative to solar composition (Nissen et al. 1994, 
Magain 1989, Wheeler, Sneden \& Truran 1989). Oxygen is enhanced at about the 
same level as the rest of the $\alpha$--elements. This is consistent 
with analytical models for the chemical evolution of the galaxy (Pagel 
\& Tautvaisiene 1995).

Therefore, there is strong observational evidence that the $\alpha$--
elements are enhanced relative to Fe in the GCs. An important point then 
is: do we need new evolutionary tracks? 
The subject is still an open question. Two groups of researchers have 
 put forward answers to the problem without performing detailed calculations of 
stellar evolution with an arbitrary abundance of $\alpha$--elements.

Salaris, Chieffi and Straniero 
(1993) have studied the problem by comparing series of models with  
special combinations of  enhanced $\alpha$--element abundances for 
low mass stars. They 
concluded that the effect of $\alpha$--element enhancements is 
well simulated by scaling the metallicity using the formula: 

\begin{equation}
Z=Z_0(0.638f_{\alpha}+0.362)
\end{equation}

where $Z_0$ is the metallicity scaled according to iron abundance and 
$f_{\alpha}$ is the enhancement factor. From data in the literature a 
typical value for $f_{\alpha}$ is $\sim$ 2. 
A main concern with this procedure is that the effect on the opacities 
from the $\alpha$--elements has not been `really' calculated. The 
basic assumption by Chieffi et al. is that opacity scales as the number 
of $\alpha$--elements. It is clear that simply scaling the Rosseland 
mean opacity cannot be the correct approach, and that a definitive 
answer will have to wait for the availability of opacities for any 
arbitrary composition.  

The argument from the group by VandenBerg and collaborators against 
Chieffi et al.'s approach is that oxygen does not contribute at very 
low metallicity to the interior opacities. The reason for this is that 
at these low metallicities the opacity source is mainly due to free--free 
transitions of electrons from H and He. The authors argue that the rest of the 
$\alpha$--elements do not contribute to the interior opacities, only 
to the boundary conditions in the stellar atmosphere. Therefore the 
only effect of the $\alpha$--elements would be through the catalytic 
effect of oxygen in the CNO cycle and not in the opacities. Following 
this argument there is no need for arbitrary scaled opacities. 
 
Recognising that a detailed and realistic calculation,  where all the
opacities for the various abundances are included, is the only way to
test the effects of non- solar abundances on stellar evolutionary
tracks, we decided to use both the approaches by Chieffi et al. and by
Vandenberg  to compute masses and ages for the set of GCs.  Also we
compared with simple solar scaled models i.e. log$ Z_0=$ log$ Z_{\odot}
+$ [Fe/H].

It is important to know how well our parameterisation by $\alpha$ and 
$\eta$ approximates reality. Obviously these two parameters represent 
only an empirical approximation to the real physics. They represent a 
parameterised macroscopic description of complex phenomena -- convection 
and mass loss 
-- that are happening in real stars. The point then to understand the 
role of $\alpha$ and $\eta$ is to link both of them to observations. 
This requires that a sample of statistically useful data be gathered for 
them. In the case of $\alpha$, a good number of stars is found at any 
point of the HR--diagram, but it is important to select one area where 
the physical conditions of the stars, in particular $M$, are the same. 
The RGB is particularly useful because the mass is strongly constrained 
to almost one single value along the upper part of the RGB.  
The $\eta$ parameter has to be treated in the same way as $\alpha$. In 
this case the mass loss is only important at the tip of the RGB.
($\dot M = 1.27 \times 10^{-5} \eta M^{-1} L^{1.5} T_{\rm eff}^{-2}$).

\section{HB morphology from variations in the mass loss efficiency}

The spread of stars along the HB is mainly due  to previous mass loss
which varies stochastically from one star to another (Rood 1973). The range of
colours where zero-age HB stars are found is a function of metallicity
(the ``first parameter'') and of the range of ZAHB masses.  More
precisely, the ZAHB colour at given metallicity depends on both the
star's total mass and the ratio of core mass to total mass, but the
core mass is essentially fixed by the physics of the helium flash and
is quite insensitive to the mass and metallicity.  For a given average
mass loss, the average final mass is thus a decreasing function of age,
which is therefore a popular candidate for the ``second parameter''
(Searle \& Zinn 1978), although other candidates such as CNO abundance
have also been suggested. A strong case for age as the chief (though
perhaps not necessarily the only) second parameter has been made by
Lee, Demarque \& Zinn (1994), who find a tendency for the clusters to
be younger in the outer Galactic halo.  J\o rgensen \& Thejll (1993),
using analytical fits to a variety of RGB models and following
evolution along the RGB with mass loss treated by Reimers's (1975)
formula, showed that, for clusters with narrow RGB's (the majority),
star-to-star variations in initial mass, metallicity or mixing-length
parameter can be ruled out as a source of the spread along the HB,
leaving as  likely alternatives only either variations in the Reimers
efficiency parameter $\eta$ (or some equivalent) or a delayed helium
flash caused by differential internal rotation. The latter alternative
would lead to a fuzzy distribution of stars at the RGB tip.

With our data we can analyse these propositions. Assume that there was a 
variation in the total mass at the flash, caused by mass loss. Looking 
at Fig. 10 we see that the effect on the luminosity at  the helium core 
flash is small $\sim 0.01$ mag, but the effect on the temperature is 
quite significant $\sim 110 K$. On the other hand a delayed helium core 
flash would not produce any effect on the effective temperature but 
would make stars appear above the theoretical helium core flash in a bin 
of $\sim 0.3$ mag. Considering that the evolution time in this very last 
bin would be the same as in the last bin before the theoretical helium core 
flash, we would expect the same number of stars in these two bins  of the 
diagram. So, for a typical GC we would expect 3--4 stars. Variations in 
the mass loss will certainly produce variations in the morphology  at 
the RGT. From Fig. 10, we can see what the effect on the position of the 
RGT is depending on the different values for $\eta$. The `bending' of 
the RGB to lower temperatures should be, therefore, observed in 
HR--diagrams from GCs. This `bending' should also put a constraint 
on the mass distribution of stars on the HB but deficiencies with 
precise atmospheric boundary conditions (Alexander 1994), make this 
evidence qualitative rather than quantitative. 

Following this strategy we looked at the previous set of observations
and counted the number of stars that were expected in every bin of
luminosity.  Using the set of three clusters where was possible to
distinguish the RGB from the AGB (M72, M68 and M5), we had a relatively
good statistical sample to test the theory of a delayed helium core
flash.  We counted the RGB stars and compared them with the theoretical
predictions.  In order to calculate the number of stars expected in
every bin of luminosity we used the SSE to compute the time spent there
and then used the fuel consumption theorem (Renzini \& Buzzoni 1986) to
compute the number of stars -- the integrated luminosity of the cluster
was properly scaled to the area covered by the CCD.  We have concluded
from the set of observations that there is no GC where there appears to
be an extra number of stars populating the RGB beyond the helium core
flash (Fig. 11 -- Fig. 14).  This argument rules out, to a level of
0.01 \Msun, variations of the core mass at the flash as the cause of HB
colour variations.

Therefore we are left with stochastic variations in the mass loss
efficiency along the RGB as the only explanation to account for the HB
morphology.  An additional possible cause for the required mass loss
could in theory  be that the core flash is affecting the structure of
the uppermost layers in such a drastic way that it could make the star
lose mass at the  RGT.  We emphasise that the typical mass lost during the
uppermost part of the RGB is about 25\% of the total stellar mass. If
such a large fraction of the star were to be lost to one violent effect
(the He core flash), then it is very unlikely that it could happen
without leaving spectral changes due to mixing of material from the
core region to the surface, which are not seen in the subsequent HB
star.

 It is therefore meaningful to proceed to an analysis of both the RGT
and the HB and link them together to deduce general properties from
morphological arguments.

The SSE is the tool that we use to model the evolution along the RGB -- 
including mass loss -- and calculate the properties of the stars on the 
HB. In addition to this we perform some more steps to fully analyse the 
physical parameters on the RGB and HB of GCs.   The procedure that we use 
in combination with the SSE to analyse the morphology of the RGB and the 
HB together and constrain the mass of the stars at the RGB proceeds in 
the following way: 
 
\begin{itemize}

\item The mass on the upper part of the RGB is determined from
 the average mass of the HB, the average mass loss efficiency and its
dispersion. Calculating the average mass and then the 2$\sigma$ value
of the distribution will give the range of masses along the HB. The
mass difference between stars along the HB is less model dependent than
the individual mass determination star by star and the average mass of
the HB is better determined than the individual masses of the stars
along it. The argument is twofold: the first obvious reason is that the
number of stars increases and the error is statistically reduced, the
second reason is that using `canonical' coordinates like the ones used
in Crocker, Rood and O'Connell (1988) for the zero age horizontal
branch location (ZAHB) the uncertainty in the mass reduces to the
uncertainty in the choice of $Z$ but not in the location of the stars
on the ZAHB (Crocker, Rood and O'Connell 1988).  Apart from this, the
individual masses of the stars are much less grid dependent than the
mass on the RGB. This is because $M_{c}$ is very well constrained to a
narrow range from theory (Jimenez et al. 1995) and therefore the
luminosity as well.  Also, the model grid dependence is very low. A
comparison between the grid by Sweigart (1987) and the one by Dorman
(1992a) gives a difference between  models with identical total mass
and core mass of 0.02 dex in log\,T$_{\rm eff}$, and 0.01 dex in
log\,$L/L_{\odot}$. This implies a difference of 0.01 $M_{\odot}$ in
the  total mass at the ZAHB. Using a procedure similar to the one
described in Crocker, Rood and O'Connell (1988) for positioning stars
on the ZAHB we have determined the average mass at the HB for the set
of GCs.  Knowing this value and the statistical spread in mass around
it, we can calculate the mass of the reddest part of the HB. The
reddest part of the HB will correspond in practice to the 2$\sigma$
value of the statistical distribution. This will give us the $M_{RGB}$,
since stars with no mass-loss along the RGB will fall in the reddest
part of the HB and we find that $<\eta>-2\sigma = 0$. Then the value of
$\eta$ can be computed since it has to reproduce the mean mass of the
HB, constrained by the fact that $\eta=0$ has to reproduce the
2$\sigma$ value for the mass of the HB at its reddest part.

\item A fit is found for the best value of $\alpha$ using the RGB. As
we have shown the RGB is sensitive to the value of $\alpha$. It is
found that the same  value of $\alpha$ fitted all the RGBs (Table 3,
J\o rgensen \& Thejll 1993).

\item The range in colour of the HB is reproduced by including mass
loss along the RGB.  In this way the mass of the RGB is strongly
constrained since it has to reproduce the morphology of the HB.

\item Now we have all the necessary parameters to model the RGB and the 
HB. With these data we can calculate a track and give the age at the RGT, 
and therefore the age of the GC itself.  

\end{itemize}

In Fig. 15 we show the results of fitting tracks to the RGB
 with a value of $\eta = 0.0$. The distance between the abscissa of the
track at the RGT and the observed stars will give an additional
estimate of $<\eta>$. As we have discussed before, mass loss does not
affect the base of the RGB but it does affect the morphology of the
RGT.

The importance of this procedure is that it gives a statistical map of 
the distribution of the mass in the HB and it sets limits to the value of 
$\eta$. The results for the nine clusters show that the value of $\eta$ 
does not exceed 1.0, and that the average value is 0.4. 
This is in good agreement with observations on field 
giants (Kudritzki \& Reimers 1978, Wood \& Cahn 1977). 

\begin{figure}
\centering
\leavevmode
\epsfxsize=1.0
\columnwidth
\epsfbox{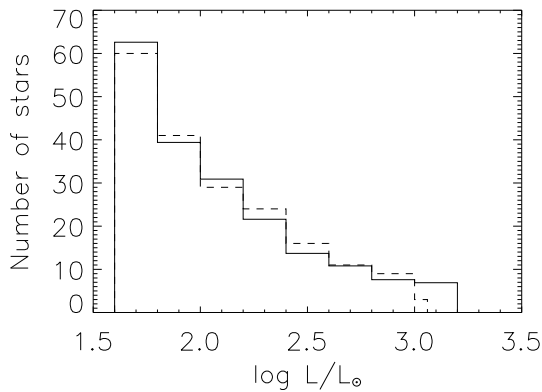}
\caption[]{The histogram shows the number of stars that populate the
RGB for M68. The continuous line shows the theoretical predictions and
the broken line the observations.
 We have calculated the number of expected stars in every bin of
luminosity using the formula in Renzini \& Buzzoni (1986); to do it we
have used the actual field of the CCD. } 
\end{figure}


\begin{figure}
\centering
\leavevmode
\epsfxsize=1.0
\columnwidth
\epsfbox{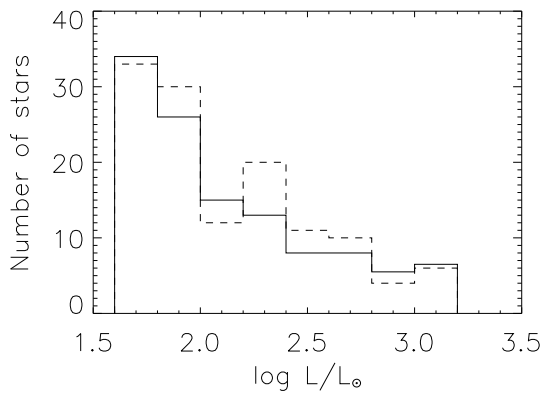}
\caption[]{For the cluster M5 due to the good split between the RGB and 
the AGB it was easy to distinguish if there was any `delayed' RGB star 
above the theoretical helium core flash point. None was found, in perfect 
agreement with the theory. The histogram confirms this result.}
\end{figure}

\begin{figure}
\centering
\leavevmode
\epsfxsize=1.0
\columnwidth
\epsfbox{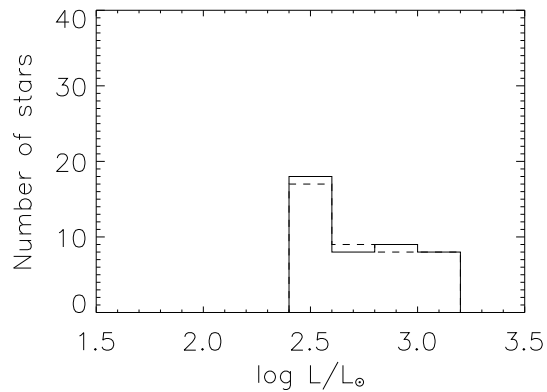}
\caption[]{For M72 the agreement between the theoretical number of 
stars per luminosity bin and the observed ones is very good. The dashed 
line merges with the solid line for the last luminosity bin.}
\end{figure}


\begin{figure}
\centering
\leavevmode
\epsfxsize=1.0
\columnwidth
\epsfbox{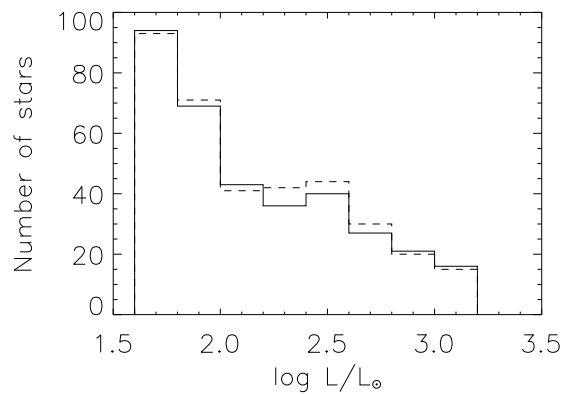}
\caption[]{The figure shows the combined RGB histogram for the three GCs 
where it is possible to distinguish the RGB from the AGB -- M68, M5 and M72. 
This allows to increase the statistics close to the RGT.
The agreement between theory and observations is very good. There are 
no stars beyond the theoretical helium core flash.  
}
\end{figure}

\section{Analysis of the HB in the GCs}

Using the above method, we have analysed the RGB and HB of eight GCs.
For the whole set of GCs the procedure has been the same. The first
step was to compute the mean HB mass using the procedure described in
Crocker, Rood and O'Connell (1988). This procedure locates the set of
HB stars in a GC in a `canonical' set of coordinates -- in the
luminosity--$T_{\rm eff}$ diagram -- in order to reduce the model
dependence of the mass determination. Once we know how the stars of the
HB are distributed in the luminosity--$T_{\rm eff}$ diagram (we have
used Kurucz (1993) model atmospheres), we can compute their masses
using different grids of HB models. As mentioned before we have used
three different approaches to account for the $\alpha$--elements. For
the case where only oxygen was enhanced we have used Dorman (1992b)
models to compute the masses, results are presented in Table 3 column
6.  In order to follow Chieffi et al.'s approach we have used
Castellani, Chieffi \& Pulone (1991) models with Z given by equation 1
and $f_{\alpha}$=2, the results are marked in Table 3 in column 8. The
same set of models has been used for simple solar--scaled metallicities
and is given in Table 3 column 7.  Once the mean mass of the HB is
determined and the proper value of $\eta$ is applied, it is
straightforward to determine the age of the GC, since now we know the
mass of the RGB stars. We have used our grid of models (Jimenez \&
MacDonald 1995) to compute the ages of the cases with $Z_{0}$ and $Z$,
and the models by Bergbusch \& VandenBerg (1992), to compute the ages of
the O-enhanced case. The various physical parameters determined for the
set of GCs are shown in Table 3. The uncertainty in all age
determinations is 2 Gyr.

\begin{figure*}
\centering
\leavevmode
\epsfysize=13.0cm
\epsfbox{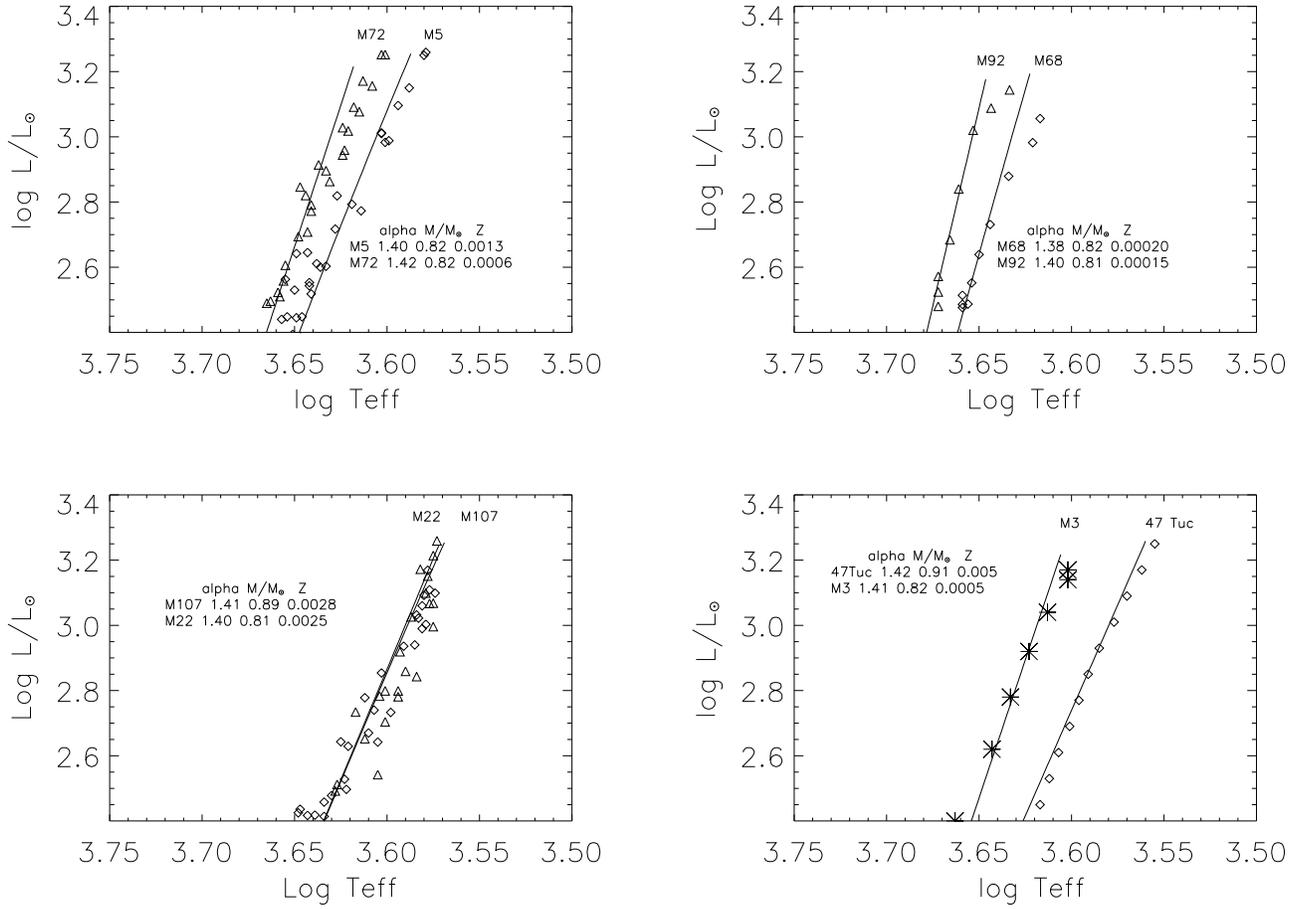}
\caption[]{The RGBs of the GCs fitted in this paper}
\end{figure*}

\begin{table*}
\begin{center}
\begin{tabular}{cccccccccccccc}
1&2&3&4&5&6&7&8&9&10&11&12&13&14\\ 
&$Z_{0}$&$Z$&$\alpha$&$<\eta>$&$M_{HB}{\it 1}$&$M_{HB}{\it 2}$&$M_{HB}{\it }3$&$M_{\rm RGB}{\it 1}$&$M_{\rm RGB}{\it 2}$&$M_{\rm RGB}{\it 3}$&t{\it 1}&t{\it 2}&t{\it 3}\\
& & & & & & \\
&&&&&O-enh&$Z_{0}$&$Z$&O-enh&$Z_{0}$&$Z$&O-enh&$Z_{0}$&$Z$\\
\hline
M92&0.00015&0.0002&1.40&0.4&0.67&0.71&0.71&0.77&0.81&0.81&12.9&13.2&13.2 \\
M68&0.0002&0.0003&1.38&0.4&0.68&0.73&0.73&0.78&0.82&0.82&12.6&12.6&12.7 \\
M22&0.0004&0.0007&1.40&0.4&0.64&0.69&0.69&0.78&0.81&0.81&13.5&13.5&13.7\\
M3&0.0005&0.0008&1.41&0.4&0.65&0.72&0.72&0.78&0.82&0.82&13.2&13.0&13.5\\ 
M72&0.0006&0.0009&1.42&0.4&0.65&0.71&0.71&0.78&0.82&0.82&12.7&13.0&13.5 \\
M5&0.0013&0.0021&1.40&0.4&0.65&0.70&0.70&0.79&0.82&0.82&12.9&13.0&13.8 \\ 
M107&0.0028&0.0046&1.41&0.4&0.72&0.77&0.77&0.85&0.89&0.89&12.3&12.0&13.2\\
47~Tuc&0.005&0.008&1.42&0.4&0.74&0.78&0.78&0.88&0.91&0.91&11.6&11.5&13.0\\ 
\hline
\end{tabular}
\caption{The table shows the physical parameters calculated for the GCs. 
$Z_{0}$ is the solar scaled metallicity according to [Fe/H].  $Z$ 
represents the metallicity in the GC according to Salaris, Chieffi and 
Straniero (1993), but with $f_{\alpha}=2$. $\alpha$ is the mixing length 
parameter fitted from the RGB. $<\eta>$ has been calculated from the value 
to reproduce both the point in the HB with 
$\eta=0$ and the the mean average 
HB mass, also from the `bending' of the RGT -- except for M68 and M92, where 
the determination from the `bending' was not possible and only the first method 
was applied. Column 6 gives the mean HB mass calculated from the models by 
Dorman (1992b) with O/Fe enhanced . The following two columns represent 
the computation of the mean HB mass from the models by Castellani, Chieffi 
and Pulone (1991), using effective metallicities $Z_{0}$ and $Z$ respectively. 
The next three columns show the corresponding values  for the mass at the RGB; 
 {\it 1} corresponds 
to models with only [O/Fe] enhanced, {\it 2} models with metallicity $Z_0$ 
and {\it 3} models with metallicity $Z$. }
\end{center}
\end{table*}

\section{Discussion}
It is interesting to discuss what would be the age of the oldest GCs if 
all the uncertain parameters -- mass determination, metallicity, helium 
content -- are pushed in the same direction within the quoted errors. 
The uncertainty in $Y$ is only $\pm 0.01$ (see Pagel 1992), 
which results in an age uncertainty of 0.7 Gyr. For a cluster like M92 
with a well determined $Z$-value of 0.0002, a change of 0.0001 will 
produce a change in the age of 0.1 Gyr. As we have discussed, an 
uncertainty in the mass of 0.03 $M_{\odot}$ will produce an error on the 
age estimate of 2 Gyr. As first shown by Stringfellow et al. (1983), 
gravitational settling of helium and the concomitant displacement of the 
major nuclear fuel, hydrogen, from the stellar core has the effect of 
shortening main sequence lifetimes. Recent studies by Proffitt \& 
VandenBerg (1991) and Chaboyer et al. (1992) find an age reduction of 
about 1Gyr. It is important to note that although this process has a 
major effect on the position of the main sequence turnoff in the HR 
diagram, the effect on the position of the RGB is negligible. This means 
that our method is basically unaffected by the process of helium 
settling, except that, if helium settling is important and not washed 
out by mixing processes, our ages from standard evolutionary tracks 
would have to be reduced by 0.5--1.0 Gyr. In the most extreme case, 
combination of these effects gives a maximum age reduction of 3.8 Gyr, 
so that for M92 we would obtain an age as low as 9.7 Gyr. Therefore GCs 
ages as low as 10 Gyr cannot be totally ruled out, but it is important 
to emphasise that this is the extreme lower limit, and that unless 
stellar evolution theory is completely wrong or some hidden physics is 
playing an important role, ages of GCs cannot be lower than 10 Gyr.

It is important to notice that variations needed in $M_{c}$ alone to produce 
the observed spread in colour in the HB are about $0.1 M_{\odot}$, but 
would also produce a `vertical' spread in luminosity of 0.5 mag --
according to the models by Dorman (1992a) and Castellani, Chieffi \& Pulone 
(1991). This is 
obviously much bigger than the real spread observed in GCs' HBs (Fig. 2--
6). This is another argument to rule out the theory of a delayed core 
flash, and also rules out the possibility that random variations in the 
core mass due to the helium core flash take place between the RGT and the HB 
rapid evolution phase.   

Those GCs that present a thick RGB show also a spread in luminosity on 
the HB. Using the models by Dorman (1992a) and Castellani, Chieffi \& Pulone 
(1991) we see that 
a spread of 0.5 dex in [Fe/H] would produce a spread in the luminosity of 
the HB of 0.3 mag. It is nice to see that this is roughly the spread 
observed in the HB of M22 and M107. On the other hand clusters with a 
very well defined and thin RGB like M68 show an admirably thin HB. It is 
tentatively concluded that the spread in luminosity on the HB could be 
due to different metallicities, but two clusters give too little statistics 
to draw a definitive conclusion. A large number of clusters has to be 
observed. 

In the calculation of the age of the globular clusters the main
ingredient and delicate point is the mass determination. An error in
the mass of $0.05 M_{\odot}$ will lead to an uncertainty in the age of
3 Gyr.  Therefore it is very important to know the accuracy of our mass
determination. From the different grids of models published we found a
spread of $0.01 M_{\odot}$. The procedure to calculate the average mass
of the HB is very consistent in itself and the internal accuracy of the
procedure is $0.005 M_{\odot}$, which is smaller than likely
systematic effects related to uncertainties in the physics of the horizontal
branch (semi-convection etc.) and the effects of non-standard chemical
composition.
An error of $\pm 0.03 M_{\odot}$ will give an uncertainty of $\pm 2$ Gyr.
(Jimenez \& MacDonald 1995), which we believe to be a reasonable estimate of
our uncertainties. One advantage of our method is that it is virtually 
independent of the distance modulus. In 
comparison, the age determination from isochrone fitting has a typical
 uncertainty of 3--4 Gyr. as we have shown from a bad choice of the
 mixing length and a bad definition of the turnoff itself, as well as
 uncertainties in the distance. 

From the previous photometric data for the RGBs of the set of
GCs, it was clear that the non-existence of `delayed' RGB stars gives
an obvious method to calculate the distance modulus of the cluster.
Since the luminosity of the RGT does not change much in the mass range
0.8--0.9 $M_{\odot}$ -- 0.04 mag (Jimenez \& MacDonald 1995) -- and is
almost independent of metallicity, a fit to the observed histogram at
the last leg of the RGB would give a very accurate determination of the
distance modulus. We have used this procedure to make a consistency
test of our method and recalculate the distance modulus. In Table 4 we
present the results of our fitting procedure to the RGT with
theoretical models. As we pointed out before the intrinsic error in
this determination is 0.04 mag.

\begin{table}
\begin{center}
\begin{tabular}{ccc}
 & $(m-M)_v$ fit to RGT & $(m-M)_v$ prev. \\ 
\hline
M68 & 15.20 & 15.25 \\
M22 & 13.60 & 13.50 \\
M72 & 16.30 & 16.50 \\
M5 & 14.51 & 14.53 \\
M107 & 15.03 & 14.97 \\
M3 & 15.00 & 15.01 \\
M92 & 14.50 & 14.45 \\
47 Tuc & 13.46 & 13.46 \\
\hline
\end{tabular}
\end{center}
\caption{The third column shows the values of the distance modulus 
for the set of GCs observed in this study. The distance moduli have been 
calculated from the fit to the upper part of the RGB up to the RGT. This method 
gives a general uncertainty of 0.05 mag -- see explanation in the text. The 
column to the right represents the values adopted from the literature.}
\end{table}

Assuming the masses we have derived to be correct,
an important question is how much of the missing physics could affect the 
age determination in our method. Two scenarios that could seriously 
affect the method  are related to helium diffusion and a more realistic 
treatment of the opacity problem for non--scaled solar abundances. 
The problem of the opacities for the $\alpha$--elements has been 
discussed in the text and there is much neglected in this field. A 
necessity for arbitrary composition stellar evolution sequences is 
obvious. 

The helium diffusion problem has been studied by  Proffitt \& VandenBerg 
(1991) and Chaboyer et al. (1992). They find that it will lead to an age 
reduction of about 1Gyr. The reason for this age reduction comes from the 
fact that less H is available for burning due to He sinking into the 
core. 
 It is important to notice that the position of the RGB and the HB are 
almost unaffected by this process -- while the position of the turn--off 
point obviously is. This means that our method is basically affected by 
the process of helium diffusion only to the extent that this affects the 
evolutionary lifetime along the MS. As a result of this, if helium 
diffusion is proven to be important,   
 we would have to cut by 0.5--1.0 Gyr. the ages calculated 
from our standard evolutionary tracks.        

The `second parameter' problem refers to clusters with the same 
(intermediate) metallicity but different HB morphologies. 
 The most common 
resource is to explain it by age differences among the clusters.
>From our study we have concluded that the origin of the HB morphology 
is due to a spread of the mass loss efficiency along  the 
RGB, but centred around a well defined value of $\eta=0.4$. 
This may or may not be the case for the `second-parameter' clusters. However, 
our results show that, if indeed these clusters are younger, then their 
initial mass of the stars on the RGB will be slightly greater, and then a 
similar mass loss rate superposed on an unchanged core mass will lead to thicker envelopes and a redder HB. Therefore, 
a consistency result of the method claims that since the average $\eta$ 
is the same for all the clusters, the mass at the RGB has to be different 
in order to produce different HB morphologies. This would imply that 
age difference is, in fact, the explanation for the `second parameter' 
problem. We therefore confirm previous solutions to this problem (Lee, Demarque 
\& Zinn 1994). All studies that include the HB get a good agreement on the age, 
e.g. 47 Tuc (Dorman, VandenBerg \& Laskarides 1989).

Finally, we make some comments on the comparison of our results with
those of the more conventional method based on turn-off luminosity,
itself depending on the magnitude difference $\Delta V$ between the HB and
the turnoff.  The HB is calibrated either on the basis of HB
models, such as we have also used (e.g. Chaboyer, Sarajedini \& Demarque
1992; Salaris, Chieffi \& Straniero 1993), or from luminosities of RR Lyrae
stars based on Baade-Wesselink pulsation analysis (Carney, Storm \& Jones
1992), the same adjusted to fit extragalactic cepheids (Walker 1992), or
analysis of the Oosterhoff period-shift effect (Sandage 1993).  The
Walker and Sandage scales give the greatest distances and hence the
shortest ages (the range of about 0.3$^m$ between HB calibrations gives a
range of 25 per cent in age, while uncertainty in $\Delta V$ itself gives
a further 10 per cent or so and more in some cases; cf. comments by Carney
et al. on M68, to which they assign  an age of 21.3 Gyr (taking $[\alpha/H]
= [O/H] = 0.3$), the same as for M92, although the formal result from $\Delta V$
is only 16.4 Gyr). Bergbusch \& VandenBerg (1992), using oxygen-enhanced
models, suggest an age of 14 Gyr for M92, very similar to our values; this
requires adoption of a relatively large distance modulus, (m-M)$_V$ = 14.7,
compared to Carney et al.'s adopted modulus of 14.3; these moduli essentially
straddle the range between extreme (semi-empirical) calibrations of RR Lyrae
luminosities, while our adopted modulus for M92 is 14.5. The remaining
discrepancy between our value and that of Bergbusch \& VandenBerg, when
their modulus is replaced by ours, is about 3Gyr, a gap that is readily
bridged by differing model assumptions.

This last claim is supported by a comparison with the work of Chaboyer,
Sarajedini \& Demarque (1992), who use an $\alpha$-enhanced chemical
composition that seems to us very realistic ($[O/Fe]=[\alpha/Fe]=0.4$;
cf. Pagel \& Tautvaisiene 1995), and that of Salaris, Chieffi \& Straniero
(1993)  who use a somewhat more  $\alpha$-enhanced mixture. A comparison of
the clusters that we have in common is given in Table 5.

\begin{table*}
\begin{center}
\begin{tabular}{ccccccc}
& (m-M)$_{V}$ Chaboyer & Age (Gyr) & (m-M)$_{V}$ Salaris & Age (Gyr) & (m-M)$_{V}$ This work & Age (Gyr) \\
\hline
M92 & 14.6 & 17.0 & 14.5 & 18.4 & 14.5 & 13.2 \\
M68 & 15.2 & 12.9 & 15.15 & 14.2 & 15.2 & 12.7 \\
M3 & 15.1 & 14.2 & 15.0 & 15.6 & 15.0 & 13.5 \\
M5 & 14.5 & 13.3 & 14.4 & 15.0 & 14.5 & 13.8 \\
M107 & 15.0 & 14.0 & 14.85 & 15.9 & 15.0 & 13.2 \\
47 Tuc & 13.35 & 14.0 & 13.2 & 18.0 & 13.5 & 13.0 \\  
\hline
\end{tabular}
\end{center}
\caption{The table shows the different age estimates for 
different assumptions of the distance modulus. Also, we show 
the comparison of conventional main sequence turn-off fitting 
with our method.} 
\end{table*}

It transpires from the table that our ages are not in serious
disagreement with those deduced from turn-off magnitudes, bearing in
mind differences in adopted distance, and the large discrepancies that
occasionally occur even when the same distance is adopted.  Chaboyer
(1995) quotes an average age of 14.2 Gyr for the lower-metallicity
clusters, using the Walker distance scale, and a minimum possible age of
11 Gyr.  There is thus no evidence
for serious systematic errors in our method, and we consider that the
fit to the RGB luminosity function that we have made provides a more
robust method of distance determination than the RR Lyrae method.  Our
results also show that internal rotation and diffusion effects have little
influence on the HB core masses.

\section{Conclusions}

In this paper we have presented clues for two important questions in 
stellar evolution and cosmology: the spread in colour of the HB and the 
age of the oldest known stars in the universe. Using very accurate 
photometry that we have obtained on five globular clusters we were able 
to distinguish for three of them the RGB from the AGB with no ambiguity. 

With these data we studied the possible scenarios to produce the spread 
in colour on the horizontal branch. The theory of a delayed helium core 
flash would produce an extra number of stars above the theoretical helium 
flash point. We have seen that this is not happening. Also, variations in 
the core mass would produce a vertical spread in the HB that is not observed, 
 except in cases where there is also a spread in the RGB -- due to 
metallicity variations -- and these metallicity variations could  
account for the vertical spread. The only scenario left to explain the 
spread in colour along the HB is that variations in the mass 
loss along the RGB produce a different ratio of total mass to core mass. 
As a consequence of this we have concluded that the explanation for 
the `second parameter' problem relies on age differences among the 
clusters that present this effect -- different masses at the RGB. 
Even though a different value for $<\eta>$ among these clusters could 
produce the same result, the question would be, why should these clusters 
have a different value of $<\eta>$?

Once the nature of the HB has been explained we have used  the morphology 
of the RGT to constrain the amount of mass that is lost at this 
stage of stellar evolution. Using this and the morphology of the HB we 
have been able to put strong constraints on the mass of the RGB stars. We 
have calculated ages for the GCs in the sample and found that the oldest 
clusters have an age of 13.5 Gyr. This estimate of the age is in better 
agreement with current cosmological models, especially an open universe 
with $H_{0} \approx 80 $km s$^{-1}$ Mpc$^{-1}$ (Freedman et al 1994, Pierce et al 1994).

Now we are in the position to answer all the questions that we 
formulated in the introduction:
\begin{itemize}
\item [i)] The HB morphology is well explained by differential 
variations in the mass loss efficiency along the RGB among the stars. 
The distribution of the HB mass is gaussian. In 
the case of $\eta$ a value of $<\eta>=0.4$ is found for the set of GCs 
studied.

\item [ii)] Since mass loss is the cause of the HB morphology, the properties of the stars at the RGT can be linked with those at the HB. This gives a 
powerful method to constrain  the mass of the RGB.

\item [iii)] The mean mass of the HB and the 2$\sigma$ value of the 
HB mass distribution have been used to determine the value of $\eta$ and 
the mass at the RGB

\item [iv)] The HB morphology is explained as variations in $\eta$, 
but with a central value of $<\eta>=0.4$ which is the same 
for all the clusters. The 
`second parameter' problem is explained as a mass difference among 
 clusters with identical metallicity, and therefore as an age difference 
among them. 

\item [v)] The age for the oldest GCs was $(13.5 \pm 2)$ Gyr. A 1$\sigma$ 
uncertainty in each of the parameters of mass and helium content combined with the effects of helium diffusion gives a lower limit for the age of the 
oldest clusters of 9.7 Gyr.  
\end{itemize}

\section*{Acknowledgements}
RJ acknowledges support form the EU under the Human and 
Capital Mobility grant 920014. PT acknowledges support from the 
Carlsberg foundation. We thank C. Flynn, H.R. Johnson, B. Jones, B. Madore, J. Ostriker, A. Renzini and D. Seckel  for a careful 
reading of 
the manuscript and helpful discussions and comments.

\end{document}